\documentclass[aps,superscriptaddress]{revtex4}
\usepackage[utf8x]{inputenc}
\usepackage{ucs}
\usepackage{amsmath}
\usepackage{amsfonts}
\usepackage{amssymb}
\usepackage{hyperref}
\usepackage{graphicx}
\usepackage{amsthm}
\usepackage{color}
\usepackage{float}
\usepackage{wrapfig}
\usepackage{epsfig}

\newtheorem*{thm*}{Theorem}

\begin{document}

\title{Entropic measures of Rydberg-like harmonic states}


\author{J.S. Dehesa}
\email[]{dehesa@ugr.es}
\affiliation{Departamento de F\'{\i}sica At\'{o}mica, Molecular y Nuclear, Universidad de Granada, Granada 18071, Spain}
\affiliation{Instituto Carlos I de F\'{\i}sica Te\'orica y Computacional, Universidad de Granada, Granada 18071, Spain}


\author{I.V. Toranzo}
\email[]{ivtoranzo@ugr.es}
\affiliation{Departamento de F\'{\i}sica At\'{o}mica, Molecular y Nuclear, Universidad de Granada, Granada 18071, Spain}
\affiliation{Instituto Carlos I de F\'{\i}sica Te\'orica y Computacional, Universidad de Granada, Granada 18071, Spain}


\author{D. Puertas-Centeno}
\email[]{vidda@correo.ugr.es}
\affiliation{Departamento de F\'{\i}sica At\'{o}mica, Molecular y Nuclear, Universidad de Granada, Granada 18071, Spain}
\affiliation{Instituto Carlos I de F\'{\i}sica Te\'orica y Computacional, Universidad de Granada, Granada 18071, Spain}

\begin{abstract}
The Shannon entropy, the desequilibrium and their generalizations (Rényi and Tsallis entropies) of the three-dimensional single-particle systems in a spherically-symmetric potential $V(r)$ can be decomposed into angular and radial parts. The radial part depends on the analytical form of the potential, but the angular part does not. In this paper we first calculate the angular entropy of any central potential by means of two analytical procedures. Then, we explicitly find the dominant term of the radial entropy for the highly energetic (i.e., Rydberg) stationary states of the oscillator-like systems. The angular and radial contributions to these entropic measures are analytically expressed in terms of the quantum numbers which characterize the corresponding quantum states and, for the radial part, the oscillator strength. In the latter case we use some recent powerful results of the information theory of the Laguerre polynomials and spherical harmonics which control the oscillator-like wavefunctions.\\

\end{abstract}


\maketitle

\section{Introduction}

The classical and quantum entropies of the many-particle systems, which are functionals of the one-particle quantum-mechanical probability density, do not only quantify the spatial delocalization of this density in various complementary ways and describe a great deal of physical and chemical properties of the systems but also they are the fundamental variables of the information theory of quantum systems which is at the basis of the modern Quantum Information. The computational determination of these quantities is a formidable task (not yet solved, except possibly for the ground and a few lowest-lying energetic states), even for the small bunch of elementary quantum potentials which are used to approximate the mean-field potential of the physical systems \cite{sen12,batha,ghosh,aptekarev2016,dehesa2015}.\\

The harmonic oscillator is both a pervasive concept in science and technology and a fundamental building block in our system of knowledge of the physical universe \cite{bloch,moshinsky}. Indeed it has been applied from the physics of quarks to quantum cosmology. The harmonic oscillator is, together with the Coulomb potential, the most relevant quantum-mechanical potential for the description of the structure and dynamics of natural systems. It has played \textit{per se} a crucial role  in the development of quantum physics since its birth \cite{heisenberg}, mainly because the wave functions of its quantum-mechanically allowed oscillator-like states can be explicitly expressed in terms of special functions of mathematical physics (namely, the Laguerre polynomials and spherical harmonics). Moreover, it has provided an approximate model for the physically-correct quantum-mechanical potentials of many-particle systems what is very useful for the the interpretation and quantitative estimation  of numerous microscopic and macroscopic properties of natural systems. Indeed, it seems that this paradigmatic oscillator-like formalisation relies on the so-called \textit{mean-field} approximation: each particle harmonically interacts with all others in the system, regardless of their reciprocal distance. Moreover, the solutions of the wave equations of complex physical systems within this approximation are very valuable, referencial tools for checking and improving complicated numerical methods used to study such systems.\\

Let us just highlight that the oscillator wave functions saturate the most important mathematical realizations of the quantum uncertainty principle such as the Heisenberg-like \cite{sanchez,zozor} uncertainty relations, which are based on the variance and/or higher-order moments, and the entropic uncertainty relations based on the Shannon entropy \cite{bialynicki1,rudnicki}, the Rényi entropy \cite{bialynicki2,vignat} or the Fisher information \cite{romera,sanchez}. Furthermore, they have been used in numerous scientific fields ranging from quantum many-body physics \cite{yanez,bouvrie,koscik,carlos,linho1,cioslowski,armstrong2,armstrong3,armstrong4}, heat transport \cite{asadian}, quantum entanglement \cite{eisert,benavides}, Keppler systems \cite{meer}, quantum dots \cite{johnson,koscik,nazmitdinov} and cold atomic gases \cite{gajda,tempere} to fractional and quantum statistics \cite{rovenchak,schilling} and black-holes thermodynamics \cite{bombelli,srednicki}. However, the information-theoretic properties of the three (or higher)-dimensional harmonic oscillator are not yet settled down in spite of numerous efforts (see e.g., \cite{yanez_1994,batha,ghosh}), mainly because of the yet incomplete knowledge of the information theory of orthogonal polynomials and spherical harmonics \cite{yanez1999,dehesa_2001,dehesa2007,aptekarev2010}.\\

 In this work we first realize that the information entropies (Shannon, Rényi and Tsallis) of the three-dimensional single-particle systems in a spherically-symmetric potential $V(r)$ can be decomposed into two angular and radial parts. The radial part depends on the analytical form of the potential, but the angular part does not. Then, we determine both the angular contribution to these entropies for all the quantum-mechanically allowed states of the central potential $V(r)$ and the radial entropy of the highly-energetic (i.e., Rydberg) states of the (three-dimensional) harmonic oscillator in an analytical way. The latter is done by using some recent powerful results of the information theory of Laguerre and Gegenbauer polynomials \cite{sanchezarzo,aptekarev2016,guerrero2010,sanchez2010} (see also \cite{sri,comtet}).\\
 
 The structure of the work is the following. In section \ref{sec:2} we first describe the information entropies of an arbitrary probability density to be used; later we apply them to a single-particle system subject to a central potential $V(r)$, showing that they can be decomposed into radial and angular parts, with emphasis on the Rényi entropy from which all the remaining entropies can be analytically obtained. Then, we begin to calculate the information entropies of the oscillator-like states by collecting all the necessary data, particularly the quantum probability density which define these states. In section \ref{sec:3} we tackle the computation of the angular part of the Rényi entropy for all quantum states of any central potential by means of two different analytical procedures. In Section \ref{sec:4} we first calculate the dominant term of the radial part of the Rényi and Shannon entropies of the highly-energetic (i.e., Rydberg) oscillator-like states. Then, the total Rényi and Shannon entropies can be analytically obtained in a straightforward way, what is illustrated for some specific Rydberg oscillator-like states. In Section \ref{sec:5} we illustrate that the information entropies of the low-energy states can be calculated in a analytical, much simpler way; and, moreover, we show that the position and momentum values of the Shannon and Rényi entropies found for the states lying at the two extreme regions of the spectrum verify the position-momentum Shannon-entropy-based \cite{bialynicki1,rudnicki} and Rényi-entropy-based \cite{vignat,bialynicki2} uncertainty relations, respectively. Finally, some conclusions are given.  
 
 \section{Information entropies of quantum states}
 \label{sec:2}
 
In this section we define the basic information entropies of a probability density $\rho(\vec{r})$; namely, the Rényi and Tsallis entropies and their instances, the Shannon entropy and the disequilibrium. Then, we study these quantities for the quantum-mechanically allowed states of a physical system with a spherically-symmetric potential $V(r)$, pointing out that they can be decomposed into two angular and radial parts. Finally, we explicitly apply them to the oscillator-like states. Atomic units are used throughout the paper.\\

The \textit{pth-order Rényi entropy} $R_{p}[\rho]$ of the density $\rho(\vec{r})$ is defined \cite{renyi1} as
\begin{equation}
\label{eq:renentrop}
R_{p}[\rho] =  \frac{1}{1-p}\ln \int_{\mathbb{R}^{3}} [\rho(\vec{r})]^{p}\, d\vec{r}; \quad 0<p<\infty,\, p \neq 1,
\end{equation}
and the Tsallis entropy \cite{tsallis}, given by $T_{p}[\rho] = \frac{1}{p-1} (1- \int_{\mathbb{R}^{3}} [\rho(\vec{r})]^{p}d\vec{r} )$, can be obtained from the Rényi one by means of the expression
\begin{equation}
\label{eq:tsalren}
T_{p}[\rho] = \frac{1}{1-p}\left(e^{(1-p)R_{p}[\rho]}-1\right).
\end{equation}
These two sets of entropies globally quantify different facets of the spreading of the probability cloud all over the spatial volume where the density function $\rho(\vec{r})$ is defined. All the members of each set completely characterize the density under certain conditions \cite{romera_01,jizba_2016b}. Some of them are very relevant \textit{per se} such as e.g., the Shannon entropy (which measures the total extent of the density), $S[\rho] := - \int \rho(\vec{r}) \ln \rho(\vec{r}) d\vec{r}  = \lim_{p\rightarrow 1} R_{p}[\rho] = \lim_{p\rightarrow 1} T_{p}[\rho]$), and the disequilibrium (which quantifies the separation of the density with respect to equiprobability), $\langle\rho\rangle = \exp(-R_{2}[\rho] = 1 - T_2[\rho])$. See \cite{aczel,dehesa_88,dehesa_89,romera_01,jizba_2016b,leonenko,guerrero,jizbad,dehesa_sen12,bialynicki3} for further knowledge of these quantities. Let us just mention that the R\'enyi entropies and their associated uncertainty relations have been widely used to investigate numerous quantum-mechanical properties and phenomena of physical systems and processes \cite{bialynicki2,dehesa_sen12,bialynicki3,jizbad}, the pattern formation and Brown processes \cite{cybulski1,cybulski2}, fractality and chaotic systems \cite{back,jizba}, quantum phase transition \cite{calixto} and the quantum-classical correspondence \cite{sanchezmoreno} and quantum entanglement \cite{bovino,koscik1}.\\

The probability density $\rho(\vec{r})$ of a single-particle system subject to the central potential $V(r)$ is given by the squared modulus of the position eigenfunction $\Psi(\vec{r})$, which satisfies the Schrödinger equation
\begin{equation}\label{eq:schrodinger}
\left( -\frac{1}{2} \vec{\nabla}^{2} - V(r)
\right) \Psi \left( \vec{r} \right) = E \Psi \left(\vec{r} \right),
\end{equation}
where the position vector $\vec{r}  =  (x_1 , x_2, x_3)$ in polar spherical units is given as $(r,\theta,\phi)      \equiv
(r,\Omega)$, $\Omega\in S^2$, where $r \equiv |\vec{r}| = \sqrt{\sum_{i=1}^3 x_i^2}
\in [0  \: ;  \: +\infty)$ and with $\theta \in [0 \: ; \: \pi), \phi \in [0 \: ; \: 2\pi)$. It is well known that the eigenfunction factorizes as $\Psi_{n,l,m} (\vec{r})=\mathcal{R}_{n,l}(r)\,\, Y_{l,m}(\Omega)$, where the radial part $\mathcal{R}_{n,l}(r)$ depends on the analytical form of the potential and the angular part $Y_{l,m}(\Omega)$ is given by  the spherical harmonics defined \cite{nist} by
\begin{equation}
\label{eq:spherarm}
Y_{l,m}(\theta,\varphi)=A_{l,m} e^{im\varphi}\,\left(\sin\theta\right)^mC_{l-m}^{(m+\frac{1}{2})}\left(\text{cos}\,\theta\right)
\end{equation}
with the normalization constant is 
 \[
    A_{l,m}=\sqrt{\frac{\left(l+\frac{1}{2}\right)\left(l-m\right)!\left[\Gamma\left(m+\frac{1}{2}\right)\right]^{2}}
        {2^{1-2m}\,\pi^2\,\left(l+m\right)!}}\,,
 \]
and the symbol $C^{(\lambda)}_{n}(t)$ denotes the Gegenbauer polynomial of degree $n$ and parameter $\lambda$.\\
Then, the probability density of the quantum stationary state $(n,l,m)$ is given by  
\begin{equation}
\label{eq:denscentral}
\rho_{n,l,m}(\vec{r}) = \rho_{n,l}(r)\,\, |Y_{l,m}(\Omega)|^{2},
\end{equation}
where the radial part is the univariate function $\rho_{n,l}(r) = [\mathcal{R}_{n,l}(r)]^2$. Now we can compute the information entropies of this density. From Eqs. (\ref{eq:renentrop}) and (\ref{eq:denscentral}) we obtain that the Rényi entropies of the quantum state $(n,l,m)$ can be expressed as
\begin{equation}
\label{eq:renyihyd1}
R_{p}[\rho_{n,l,m}] = R_{p}[\rho_{n,l}]+R_{p}[Y_{l,m}],
\end{equation}
where $R_{p}[\rho_{n,l}]$ denotes the radial part
\begin{equation}
\label{eq:renyi2}
R_{p}[\rho_{n,l}] = \frac{1}{1-p}\ln \int_{0}^{\infty} [\rho_{n,l}]^{p} r^{2}\, dr,
\end{equation}
and  $R_{p}[Y_{l,m}]$ denotes the angular part
\begin{equation}
\label{eq:renyi3}
R_{p}[Y_{l,m}] = \frac{1}{1-p}\ln \Lambda_{l,m}.
\end{equation}
with
\begin{equation}
\label{eq:angpart}
\Lambda_{l,m} = \int_{\mathbb{S}^{2}} |Y_{l,m}(\theta,\phi)|^{2p}\, d\Omega.
\end{equation}
For $p=1$ and $2$ we obtain similar expressions for the disequilibrium and Shannon entropy, respectively. In particular, the Shannon entropy of the quantum state $(n,l,m)$ of any central potential is decomposed as
\begin{equation}
\label{eq:shannon}
S[\rho_{n,l,m}] = S[\rho_{n,l}] + S[Y_{l,m}],
\end{equation}
where the radial and angular parts are given by
\begin{equation}
\label{eq:limit1}
S[\rho_{n,l}]  =  \lim_{p\to 1} R_{p}[\rho_{n,l}],
\end{equation}
and 
\begin{equation}
\label{eq:limit2}
S[Y_{l,m}] = \lim_{p\to  1} R_{p}[Y_{l,m}],
\end{equation}
respectively. Note that, contrary to the radial parts, the angular parts $R_{p}[Y_{l,m}]$ and $S[Y_{l,m}]$ do not depend on the analytical form of the potential $V(r)$. Surprisingly, these angular Rényi and Shannon entropies have never been calculated up to now. We will do it in the next section \ref{sec:3}.\\

To go forward into the radial Rényi entropy $R_{p}[\rho_{n,l}]$, we will take into account the oscillator potential $V(r) = \frac{1}{2}\lambda^{2}r^{2}$ whose Schrödinger equation (\ref{eq:schrodinger}) is known (see e.g., \cite{yanez_1994,dong}) to be exactly solved, so that the energetic eigenvalues are
\begin{equation}
\label{eq:oscillenerg}
E_{n,l}= \lambda\left(2n+l+\frac{3}{2}\right),
\end{equation}
and the corresponding eigenfunctions are expressed as
\begin{eqnarray}
\label{eq:wavpos}
\Psi_{n,l,m}(\vec{r}) &=& \left[\frac{2n!\lambda^{l+\frac{3}{2}}}{\Gamma(n+l+\frac{3}{2})} \right]^{\frac{1}{2}}r^{l}e^{-\frac{\lambda\,r^{2}}{2}}L^{(l+1/2)}_{n}(\lambda\, r^{2}) \nonumber \\
& & \times \, Y_{l,m}(\Omega),
\end{eqnarray}
with $(n=0,1,2,\ldots; l=0,1,2,\ldots; m=-l,-l+1,...,+l)$, and $L^{(\alpha)}_{n}(t)$ denotes \cite{nist} the Laguerre polynomial of paramater $\alpha$ and degree $n$.\\

Then, the position probability density of the isotropic harmonic oscillator has the form (\ref{eq:denscentral}) where the radial part is given by
\begin{eqnarray}
\label{eq:denspos}
\rho_{n,l}(r) &=& \frac{2n!\lambda^{l+\frac{3}{2}}}{\Gamma(n+l+\frac{3}{2})}r^{2l}e^{-\lambda\, r^{2}}\left[L^{(l+1/2)}_{n}(\lambda\, r^{2})\right]^{2}\nonumber \nonumber \\
&=& \frac{2n!\lambda^{\frac{3}{2}}}{\Gamma(n+l+\frac{3}{2})}x^{\frac{1}{2}}\omega_{l+\frac{1}{2}}(x)\left[L^{(l+1/2)}_{n}(x)\right]^{2}\nonumber \\
&=& 2\,\lambda^{\frac{3}{2}}\frac{\omega_{l+\frac{1}{2}}(x)}{x^{-1/2}}[\widehat{L}_{n}^{(l+1/2)}(x)]^{2} 
\end{eqnarray}
where $x=\lambda\,r^{2}$ and
\begin{equation} \label{eq:c1.1}
\omega_{\alpha}(x) =x^{\alpha}e^{-x}, \, \alpha=l+\frac{1}{2},	
\end{equation}
  is the weight function of the orthogonal and orthonormal Laguerre polynomials of degree $n$ and parameter $\alpha$, here denoted by $L_{n}^{(\alpha)}(x)$ and $\widehat{L}_{n}^{(\alpha)}(x)$, respectively. 
 Moreover, it is known \cite{yanez_1994} that the probability density in momentum space (i.e., the squared modulus of the Fourier transform of the position eigenfunction) is given by $\gamma(\vec{p}) =\frac{1}{\lambda^{3}}\rho\left(\frac{\vec{p}}{\lambda}\right)$. So, the position and momentum information entropies of the oscillator-like states have aformal expression of similar type.\\
 
 Later on, in Section \ref{sec:4}, we will determine in an analytical way the radial Rényi and Shannon entropies not for all quantum oscillator-like states (what is an open problem) but \textit{only} for all highly energetic (i.e., Rydberg) oscillator-like states. The Rydberg case is even a serious computational task because it involves the numerical evaluation of the Rényi and Shannon functionals of Laguerre polynomials $L_n(x)$ with a high and very high degree $n$. Indeed, a naive use of quadratures to tackle this problem is not convenient: since all the zeros of $L_n(x)$ belong to the interval of orthogonality, the increasing number of integrable singularities spoil any attempt to achieve reasonable accuracy even for rather small $n$ \cite{buyarov}. Finally, let us advance here that the information entropies for the low-energy states can be easily obtained because then the corresponding Laguerre polynomials have low degrees so that the associated entropic integrals can be solved in a analytically simple manner, as it is illustrated in Section \ref{sec:5}.
 
\section{Angular entropies of quantum states of any central potential}
\label{sec:3}
In this section we describe two qualitatively different analytical procedures for the evaluation of the angular Rényi entropy $R_{p}[Y_{l,m}]$, given by \eqref{eq:renyi3}, of any quantum state of an arbitrary central potential. Then, the corresponding angular Shannon entropies follow in the limit $p\rightarrow 1$. We start with Eqs. \eqref{eq:angpart} and \eqref{eq:spherarm} to obtain the angular functional
\begin{eqnarray}
\Lambda_{l,m} &=& \int_{\mathbb{S}^{2}} |Y_{l,m}(\theta,\phi)|^{2p}\, d\Omega\\
&=&  2\pi [A_{l,m}]^{2p}\int_{0}^{\pi} |C_{l-m}^{(m+1/2)}(\cos \theta)|^{2p}(\sin\theta)^{2pm+1}\, d\theta. \nonumber \\
&=& 2\pi [A_{l,m}]^{2p}\int_{-1}^{1} |C_{l-m}^{(m+1/2)}(t)|^{2p}(1-t^{2})^{mp}\, dt
\label{eq:angint1}
\end{eqnarray}
To compute \eqref{eq:angint1} we propose the two following methods. One based on the linearization technique of Srivastava \cite{sri,sanchezarzo} and another one based on the power expansion via the Bell polynomials \cite{sanchez2010,comtet}.

\subsection{Linearization-based method}

The functional of Gegenbauer polynomials of \eqref{eq:angint1}  can be solved by means of the linearization formula of Srivastava \cite{sri,sanchezarzo} for the natural powers of Jacobi polynomials. Indeed, since the Gegenbauer polynomials are particular instances of Jacobi polynomials as indicated by
\begin{equation}
\label{eq:gegjaco}
C_{n}^{(\lambda)}(t) = \frac{(2\lambda)_{n}}{\left(\lambda+\frac{1}{2}\right)_{n}}P_{n}^{(\lambda-\frac{1}{2},\lambda-\frac{1}{2})}(t),
\end{equation}
 where $\lambda=m+1/2$ and $n=l-m$.
we have that the angular functional $\Lambda_{l,m}$ can be rewritten as
\begin{equation}
\label{eq:angint4}
\Lambda_{l,m} = A'_{l,m}\int_{-1}^{1} |P_{l-m}^{(m,m)}(t)|^{2p}(1-t)^{mp}(1+t)^{mp}\, dt,
\end{equation}
where
\begin{equation}
A'_{l,m} = \frac{2^{2 (m-1) p+1}(2 l+1)^p}{\pi ^{2 p-1}}\left[\frac{\Gamma \left(m+\frac{1}{2}\right)^{2} (l-m)! (m!)^{2}(l+m)!}{ (l!)^{2} [(2m)!]^{2}}\right]^{p}
\end{equation}
Then, the Srivastava linearization formula appropriately modified for our purposes gives \cite{sanchezarzo}
\begin{equation}
\label{eq:lf1}
\left[ P_{l-m}^{(m,m)}(t) \right]^{2p} =\sum_{i=0}^{\infty}\tilde{c}_{i}(p,l,m) P_{i}^{(pm,pm)}(t), 
\end{equation}
(which holds for positive integer and half-integer values of the parameter $p$), where the coefficients $\tilde{c}_{i}(p,l,m)$ (or equivalently $\bar{c}_{i}(0,2p,l-m,m,m,pm,pm)$ in the notation of \cite{sanchezarzo})  have the expression
\begin{eqnarray}
\label{eq:lf2}
\tilde{c}_{i}(p,l,m) = \binom{l}{l-m}^{2p}\frac{2mp+2i+1}{2mp+i+1}  \sum_{j_{1},\ldots,j_{2p}=0}^{l-m}\sum_{j_{2p+1}=0}^{i}\frac{(mp+1)_{j_{1}+\ldots+j_{2p}+j_{2p+1}}}{(2mp+i+2)_{j_{1}+\ldots+j_{2p}}}& & \nonumber \\
\hspace{-1cm} \times \frac{(m-l)_{j_{1}}(l+m+1)_{j_{1}}\cdots(m-l)_{j_{2p}}(l+m+1)_{j_{2p}}(-i)_{j_{2p+1}}}{(m+1)_{j_{1}}\cdots(m+1)_{j_{2p}}(pm+1)_{j_{2p+1}} j_{1}!\cdots j_{2p}!j_{2p+1}! }. & &\nonumber\\
\end{eqnarray}
Substituting \eqref{eq:lf1} into \eqref{eq:angint4} and using the orthogonality property of the Jacobi polynomials 
\begin{equation}
\int_{-1}^{1} (1-t)^{a}(1+t)^{b} P_{n}^{(a,b)}(t)P_{m}^{(a,b)}(t)\, dt =
\frac{2^{a+b+1}\Gamma(a+n+1)\Gamma(b+n+1)}{n!(a+b+2n+1)\Gamma(a+b+n+1)}\delta_{m,n},
\end{equation}
we obtain the following expression for the angular functional $\Lambda_{l,m}$:
\begin{equation}
\label{eq:lf3}
\Lambda_{l,m} = A''_{l,m}\tilde{c}_{0}(p,l,m) 
\end{equation}
with
\begin{equation}
\label{eq:ados}
A''_{l,m} =\frac{2^{2p(2 m -1)+2}(2 l+1)^p}{\pi ^{2 p-1} }\frac{\Gamma (m p+1)^2}{\Gamma (2m p+2)} 
\left[\frac{\Gamma \left(m+\frac{1}{2}\right)^{2} \Gamma (m+1)^{2} \Gamma (l-m+1) \Gamma (l+m+1)}{\Gamma (2 m+1)^{2}\Gamma (l+1)^{2}}\right]^{p},
\end{equation}
and
\begin{eqnarray}
\label{eq:c0}
\tilde{c}_{0}(p,l,m) = \binom{l}{l-m}^{2p} 
 \sum_{j_{1},\ldots,j_{2p}=0}^{l-m}\frac{(mp+1)_{j_{1}+\ldots+j_{2p}}}{(2mp+2)_{j_{1}+\ldots+j_{2p}}}& & \nonumber \\
\times\frac{(m-l)_{j_{1}}(l+m+1)_{j_{1}}\cdots(m-l)_{j_{2p}}(l+m+1)_{j_{2p}}}{(m+1)_{j_{1}}\cdots(m+1)_{j_{2p}}j_{1}!\cdots j_{2p}! }. & & \nonumber \\
\end{eqnarray}
Then, taking \eqref{eq:lf3} into \eqref{eq:renyi3} one finally obtains the value  
\begin{equation}
\label{eq:angren}
R_{p}[Y_{l,m}] = \frac{1}{1-p}\ln \left[A''_{l,m}\tilde{c}_{0}(p,l,m)\right].
\end{equation}
for the angular part for the Rényi entropy of any quantum state of an arbitrary central potencial, which again hold for positive integer and half-integer values of the parameter $p$. Note that Eqs. \eqref{eq:ados} - \eqref{eq:angren} allow us to analytically compute this entropic quantity in an straightforwrad and algorithmic way.

\subsection{Bell-polynomials-based method} 

Let us now give an alternative, qualitatively different method to compute the angular Rényi functional $\Lambda_{l,m}$ given by \eqref{eq:angint1} or, equivalently, \eqref{eq:angint4}. In this method we calculate the Gegenbauer-polynomial integral involved in \eqref{eq:angint1}, or better the Jacobi-polynomial integral of \eqref{eq:angint4}, by means of the power expansion of its respective kernel. The latter is done by use of the following general result \cite{sanchez2010}: The $p$-th power of an arbitrary polynomial $y_{n}(x)$ given by
\begin{equation}
\label{eq:poly}
y_{n}(x) = \sum_{k=0}^{n}c_{k}x^{k}
\end{equation}
can be expressed as
 \begin{eqnarray}
 \label{eq:bell_poly}
 [y_{n}(x)]^{p} &=& \left(\sum_{k=0}^{n}c_{k}x^{k}\right)^{p}\nonumber \\
 &=& \sum_{k=0}^{np} A_{k,p}(c_{0},\ldots,c_{n})x^{k},
 \end{eqnarray}
where
\[
 A_{k,p}(c_{0},\ldots,c_{n}) = \frac{p!}{(k+p)!}B_{k+p,p}(c_{0},2!c_{1}, \ldots, (k+1)!c_{k}),
\]
with $c_{i}=0$ if $i>n$ and $B_{n,k}(x_{1}, x_{2}, \ldots)$ are the multivariate Bell polynomials of the second kind \cite{comtet}
\[
B_{n,k}(x_{1},x_{2},\ldots) = \sum_{\substack{j_{1}+j_{2}+\ldots=k \\ j_{1}+2j_{2}+\ldots=n}} \frac{n!}{j_{1}!j_{2}!\cdots}\left(\frac{x_{1}}{1!} \right)^{j_{1}}\left(\frac{x_{2}}{2!} \right)^{j_{2}}\cdots
\]
From the known explicit expression of the Jacobi polynomials \cite{nist} we can write
\begin{equation}
\label{eq:power_jacobi1}
\tilde{P}^{(\alpha,\beta)}_{n}(x) = \sum_{k=0}^{n}c_{k}x^{k} 
\end{equation}
with the expansion coefficients 
\begin{equation*}
c_{k} = \sqrt{\frac{\Gamma(n+\alpha+1)(2n+\alpha+\beta+1)}{n!2^{\alpha+\beta+1}\Gamma(\alpha+\beta+n+1)\Gamma(n+\beta+1)}}
 \sum_{i=k}^{n}(-1)^{i-k}\binom{n}{i}\binom{i}{k}\frac{\Gamma(\alpha+\beta+n+i+1)}{2^{i}\Gamma(\alpha+i+1)}
\end{equation*}
Then, according to \eqref{eq:bell_poly} et sequel one obtains the following expression for the $p$-th power of the orthonormal Jacobi polynomials
\begin{equation}
\label{eq:power_jacobi2}
[\tilde{P}^{(\alpha,\beta)}_{n}(x)]^{2p} = \sum_{k=0}^{2np} B_{k+2p,2p}(c_{0},2!c_{1}, \ldots, (k+1)!c_{k})x^{k} .
\end{equation}
Now, using this expression with $\alpha=\beta=m$ and $n=l-m$, and taking the resulting expression into the angular functional $\Lambda_{l,m}$ given by \eqref{eq:angint4} one has 
\begin{eqnarray}
\label{eq:ang_int}
\Lambda_{l,m} &=& \frac{\Gamma(mp+1)}{2^{p}\pi^{p-1}}
 \sum_{k=0}^{2(l-m)p}\frac{(2p)!}{(k+p)!}B_{k+2p,2p}(c_{0},2!c_{1},\ldots,(k+1)!c_{k}) \frac{[1+(-1)^{k}]\Gamma\left(\frac{k+1}{2}\right)}{\Gamma\left(\frac{1}{2}(3+k+2mp)\right)} \nonumber\\
& \equiv & \frac{\Gamma(mp+1)}{2^{p}\pi^{p-1}}\Sigma(l,m,p).
\end{eqnarray}
Thus, from \eqref{eq:renyi3} one finds the following value  
\begin{equation}
\label{eq:ren_ang}
R_{p}[Y_{l,m}] = \frac{1}{1-p}\ln  \Lambda_{l,m} \frac{1}{1-p}\ln\left[\frac{\Gamma(mp+1)}{2^{p}\pi^{p-1}}\Sigma(l,m,p)\right].
\end{equation}
for the angular Rényi entropy of an arbitrary state $(l,m)$ of any central potential $V(r)$, which holds for positive integer and half-integer values of the parameter $p$.\\

Let us finally calculate, for illustration, both the angular Rényi and Shannon entropies for some specific states by means of Eqs. \eqref{eq:ren_ang} and  \eqref{eq:limit2}.

\begin{enumerate}
\item {\bf States} $(l,l)$. From its own definition \eqref{eq:angint1} one finds the angular functional
\begin{eqnarray}
\label{eq:angint6}
\Lambda_{l,l} &=&  2\pi (A_{l,l})^{2p}\int_{0}^{\pi}(\sin\theta)^{2lp+1} \, d\theta\nonumber \\
&=&\frac{ 2^{(2 l-1) p+1} \left(l+\frac{1}{2}\right)^p}{\pi ^{2p-\frac{3}{2}}}\frac{ \Gamma \left(l+\frac{1}{2}\right)^{2 p} \Gamma (l p+1)}{\Gamma (2 l+1)^{p} \Gamma \left(l p+\frac{3}{2}\right)},\nonumber \\
\end{eqnarray}
which holds for all real values of $p$. Then, from \eqref{eq:renyi3} one finds that the angular Rényi entropy of the state $(l,l)$ is given by
\begin{equation}
\label{eq:angren7}
R_{p}[Y_{l,l}] =  \frac{1}{1-p}
\times \ln\left[\frac{ 2^{(2 l-1) p+1} \left(l+\frac{1}{2}\right)^p}{\pi ^{2p-\frac{3}{2}}}\frac{ \Gamma \left(l+\frac{1}{2}\right)^{2 p} \Gamma (l p+1)}{\Gamma (2 l+1)^{p} \Gamma \left(l p+\frac{3}{2}\right)}\right] ,
\end{equation}
Then, from this expression and \eqref{eq:limit1} one has the following value
\begin{equation}
\label{eq:angren8}
S[Y_{l,l}] = -l \left[\psi (l+1)-\psi \left(l+\frac{3}{2}\right)+\ln 4\right]
+\ln \frac{4 \pi ^2}{ 2 l+1}+\ln \frac{ \Gamma (2 l+1)}{\Gamma \left(l+\frac{1}{2}\right)^{2}},\hspace{1.5cm}
\end{equation}
for the angular Shannon entropy of the state $(l,l)$.\\

In particular, for the states $(0,0)$ and $(1,1)$ we have that
\begin{equation}
\label{eq:angint2}
\Lambda_{0,0} = (4\pi)^{1-p}\nonumber
\end{equation}
and 
\begin{equation}
\label{eq:angint5}
\Lambda_{1,1} = \frac{2^{1-3 p} 3^p \pi ^{\frac{3}{2}-p} \Gamma (p+1)}{\Gamma \left(p+\frac{3}{2}\right)},\nonumber
\end{equation}
so that the corresponding angular Rényi entropies are given by
\begin{equation}
\label{eq:angren1}
R_{p}[Y_{0,0}] =\ln(4\pi)\nonumber
\end{equation}
and
\begin{equation}
\label{eq:angren5}
R_{p}[Y_{1,1}] =  \frac{1}{1-p}\ln\left[ \frac{2^{1-3 p} 3^p \pi ^{\frac{3}{2}-p} \Gamma (p+1)}{\Gamma \left(p+\frac{3}{2}\right)}\right],\nonumber
\end{equation}
respectively, which both hold for all real values of $p$. Then, taking the limit $p\rightarrow 1$ in the two previous expressions leads us to the following values 
\begin{eqnarray}
\label{eq:angren2}
S[Y_{0,0}] &=&\ln(4\pi)\nonumber\\
\label{eq:angren6}
S[Y_{1,1}] &=& \ln \left(\frac{2 \pi}{ 3}\right) + \frac{5}{3},\nonumber
\end{eqnarray}
for the angular Shannon entropies of the states $(0,0)$ and $(1,1)$, respectively.

\item {\bf States} $(l,l-1)$. 
Operating similarly as in the previous case, one has the angular functional 
 \begin{equation*}
\varLambda_{l,l-1}=2\pi\left(\frac{(l+\frac12)(2l-1)^2\Gamma(l-\frac12)^2}{2^{3-2l}(2l-1)!\pi^2}\right)^p \frac{\Gamma(p+\frac12)\Gamma(pl-p+1)}{\Gamma(pl+\frac32)}
 \end{equation*}
 (which holds for all real values of $p$) so that the angular Rényi entropy is given by 
 $$R_p(Y_{l,l-1})=\frac1{1-p}\ln\varLambda_{l,l-1}$$ 
 and the limit $p\to1$ gives rise to the value  
 \begin{equation}
S(Y_{l,l-1})=-\ln\left(\frac{(l+\frac12)(2l-1)^2\Gamma(l-\frac12)^2}{2^{3-2l}(2l-1)!\pi^2}\right)-\psi\left(\frac{3}{2}\right)-(l-1)\,\psi(l)+l\,\psi\left(l+\frac32 \right)
 \end{equation}
for the angular Shannon entropy. For the particular case $(1,0)$, one has the angular functional
\begin{eqnarray}
\label{eq:angint3}
\Lambda_{1,0} &=&  2\pi\left(\frac{3}{4\pi}\right)^{p}\int_{0}^{\pi}|\cos\theta|^{2p} \sin\theta\, d\theta \nonumber \\
&=& \frac{3^p (4 \pi )^{1-p}}{2 p+1}\nonumber
\end{eqnarray}
and the following values
\begin{eqnarray*}
\label{eq:angren3}
R_{p}[Y_{1,0}] &=&  \frac{1}{1-p}\ln\left[ \frac{3^p (4 \pi )^{1-p}}{2 p+1}\right] ,\nonumber \\
\label{eq:angren4}
S[Y_{1,0}] &=& \frac{2}{3}+\ln \left(\frac{4 \pi }{3}\right).\nonumber
\end{eqnarray*}
for the angular Rényi and Shannon entropies.

\end{enumerate}

\section{Rényi and Shannon entropies of Rydberg-like harmonic states}
\label{sec:4}
In this section, we first determine the radial part of the position Rényi and Shannon entropies for the highly-energetic (Rydberg) states of the (three-dimensional) isotropic harmonic oscillator from their corresponding definitions \eqref{eq:renyi2} and  \eqref{eq:limit1}. Then the resulting radial values together with the angular values derived in the previous section allows us to calculate the total Rényi and Shannon entropies (as well as the Tsallis ones, because of Eq. \eqref{eq:tsalren}) of the Rydberg harmonic states, what is illustrated for some specific oscillator-like states. 

\subsection{Radial Rényi entropies}

Taking into account \eqref{eq:renyi2} and  \eqref{eq:denspos}, the radial Rényi entropy of a general oscillator-like state can be expressed as 
\begin{equation}
\label{eq:renyi4}
R_{p}[\rho_{n,l}] = \frac{1}{1-p}\ln\left[(2\lambda^{3/2})^{p-1}N_{n,l}(p) \right],
\end{equation}
where $N_{n,l}(p)$ denotes the $\mathfrak{L}_{p}$-norm of the Laguerre polynomials given by
\begin{equation}\label{eq:c1.2}
N_{n,l}(p)=\int\limits_{0}^{\infty}\left(\left[\widehat{L}_{n}^{(\alpha)}(x)\right]^{2}\,w_{\alpha}(x)\right)^{p}\,x^{\beta}\,dx\;,\quad p>0\,,
\end{equation}
where $\alpha=l+\frac{1}{2}\,,\;l=0,1,2,\ldots$ and $\beta=\frac{1}{2}(1-p)$. We note that the condition
\begin{equation}
\label{eq:condition}
\beta+p\alpha=pl+\frac{1}{2} > -1\;,
\end{equation}
guarantees the convergence of the integral (\ref{eq:c1.2}) at zero, i.e.,
 is always satisfied for physically meaningfull values of the parameters $\alpha$, $\beta$ and $p$.\\
 
Then, the problem of determination of the radial Rényi entropy of a general oscillator-like state boils down to the study of the asymptotics ($n\to\infty$) of the Laguerre norm $N_{n,l}(p)$.  The latter problem can be solved by means of the recent methodology of Aptekarev et al \cite{aptekarev2016}, which takes explicitly into account the different asymptotical representations for the Laguerre polynomials at different regions of the real half-line.\\

Moreover, this technique shows that the dominant contribution in the magnitude of the integral  comes from various regions of integration in \eqref{eq:c1.2}, which depend on the different values of the involved parameters $(\alpha,p,\beta)$. In fact, there are five asymptotical regimes which can give (depending on $\alpha,\beta$ and $p$) the dominant contribution in the asymptotics of $N_{n}(\alpha, p, \beta)$. First, at the neighborhood of zero (Bessel regime) the Laguerre polynomials can be asymptotically described by means of Bessel functions. Then, to the right of zero (in the bulk region of zeros location) the oscillatory behavior of the polynomials is asymptotically modelled via trigonometric functions (cosine regime). And at the neighborhood of the extreme right (Airy regime), the zeros asymptotics is given by Airy functions. Finally, at the extreme right of the orthogonality interval (i.e., near infinity) the polynomials have  growing asymptotics. Moreover, there are two transition regions (to be called by cosine-Bessel and cosine-Airy) where these asymptotics match each other; i.e., asymptotics of the Bessel functions for big arguments match the trigonometric function, as well as the asymptotics of the Airy functions do the same. \\

The application of Aptekarev et al' technique to the Laguerre norm \eqref{eq:c1.2} in our three-dimensional case, together with Eq. \eqref{eq:renyi4}, gives rise to the following value for the radial Rényi entropy of the Rydberg harmonic states:
\begin{equation}\label{8}
R_{p}[\rho_{n,l}]=\left\{
\begin{array}{ll}
\frac{1}{1-p}\ln\left[\lambda^{\frac{3}{2}(p-1)}C(\beta,p)\,(2n^{3})^{\frac{1-p}{2}}\,(1+\bar{\bar{o}}(1))\right],\quad & p\in(0,p^{*})\\
\\
-2\ln\left[\lambda ^{3/4}\frac{8 \sqrt{2} }{3 \pi ^{5/2}}n^{-3/4}\,(\ln n+\underline{\underline{O}}(1))\right]\,,\quad & p=p^{*}\\
\\
\frac{1}{1-p}\ln\left[(2\lambda^{\frac{3}{2}})^{p-1}C_{B}(\alpha,\beta,p)\,n^{(p-3)/2}\,(1+\bar{\bar{o}}(1))\right],\quad & p>p^{*}
\end{array}
\right.\;,
\end{equation}
with $p^{*}=\frac{3}{2}$ and the constants $C$ and $C_{B}$ are defined as
\begin{equation}\label{4}
C_{B}(\alpha,\beta,p):=2\int\limits_{0}^{\infty}t^{2\beta+1}|J_{\alpha} (2t)|^{2p}\,dt\;.
\end{equation}
for the Bessel regime, 
\begin{equation}\label{6}
C(\beta,p):=\displaystyle\frac{2^{\beta+1}}{\pi^{p+1/2}}\,\displaystyle\frac{\Gamma(\beta+1-p/2)\,\Gamma(1-p/2)\,\Gamma(p+1/2)}
{\Gamma(\beta+2-p)\,\Gamma(1+p)}\;.
\end{equation}
for the cosine regime, respectively (the symbol $J_{\alpha}(z)$ denotes the Bessel function, see e.g. \cite{nist}), and the parameters $\alpha \equiv \alpha(l)$ and $\beta\equiv \beta(p)$ are given by \eqref{eq:condition}.
\\

\textit{Hints}: To better understand the application of the previous technique to our case, let us note:
\begin{itemize}
	\item that $\beta(p^{*})-\frac{p^{*}}{2}=\frac{1}{2}-p^{*}=-1\,$, so that from (\ref{6}) we have $C(\beta,p)=\infty$. Thus, for $p\in(0,p^{*})$ the region of $\mathbb{R}_{+}$ where the Laguerre polynomials exhibit the cosine asymptotics contributes with the dominant part in the integral (\ref{eq:c1.2}). For $p=p^{*}$ the transition cosine-Bessel regime
determines the asymptotics of $N_{n,l}(p^{*})$, and for $p>p^{*}$ the Bessel regime plays the main role.

\item the $\mathfrak{L}_{p}$-norm is constant (i.e., independent of $n$) and equal to $C_{B}(\alpha,\beta,p)$, only when $(p-1)3/2-p = 0$. This means that the constancy occurs when $p=3$.  

\end{itemize}
A careful analysis of \eqref{8} shows that:

\begin{itemize}
	\item for fixed $n$ the radial entropy depends on the oscillator strength $\lambda$ in the form $-3/2 \, \ln \,\lambda$,
	\item for fixed $\lambda$ the radial entropy depends on the principal quantum number $n$ in the forms: $3/2\,\ln\,n$ (as $p\in]0,3/2[$, $+3/2\,\ln \,n - 2\, \ln \, \ln\,n$ (as $p=3/2$), constant (as $p=3$), and $\frac{p-3}{2(1-p)}\,\ln\,n$ (as $p>3$).
\end{itemize}

Since the Rényi and Tsallis entropies are related by \eqref{eq:tsalren}, the radial Tsallis entropy, $T_{p}[\rho_{n,l}]$, for the Rydberg oscillator-like states follows from \eqref{8} in a straightforward manner.

\subsection{Radial Shannon entropy}

To determine the radial part of the Shannon entropy $S[\rho_{n,l}]$ we need, according to \eqref{eq:limit1}, to compute the limit $p\rightarrow 1$ of the radial Rényi entropy $R_{p}[\rho_{n,l}]$ given by \eqref{8}. We obtain that
\begin{equation}
\label{eq:shanre}
\begin{split}
S[\rho_{n,l}] &\equiv \lim_{p\to 1} R_{p}[\rho_{n,l}] \\
& = \lim_{p\to 1}\frac{1}{1-p}\ln\Big[\lambda^{\frac{3}{2}(p-1)}C(\beta,p)\,(2n^{3})^{\frac{1-p}{2}}\,(1+\bar{\bar{o}}(1))\Big]\\
&= \left(\frac{3}{2} \ln n-\frac{3}{2} \ln\lambda +\ln \pi -1\right)\,(1+\bar{\bar{o}}(1)),
\end{split}
\end{equation}
where it can be seen that the leading term of the asymptotic expression is proportional to $\ln n$, as expected.\\

\subsection{Total position Rényi and Shannon entropies}

The total Rényi and Shannon entropies of the Rydberg harmonic states $\{n\to\infty,l,m\}$, given by Eqs. \eqref{eq:renyihyd1} and  \eqref{eq:shannon} respectively, can now be determined from the results obtained in the two previous sections in a direct, analytical and straightfroward manner. Indeed they are given by the sum of the angular part (which does not depend on $n$) obtained in section \ref{sec:3} in two different ways and the radial part which is given by Eqs. \eqref{8} and  \eqref{eq:shanre} for the Rényi and Shannon entropies, respectively. When $n$ is sufficiently large, we observe that : 
\begin{enumerate}
\item If $p\not=3$, then $\big|R_p[\rho_{n,l}]\big|>>\big|R_p[Y_{l,m}]\big|$, and so $R_p[\rho_{n,l,m}]\simeq R_p[\rho_{n,l}]\to \pm\infty$ with the same sign that $3-p$. This $n$th-asymptotical growth of the absolute value of the radial part is very very slow; the closer $p$ to 3, the slower is this growth. 
\item If $p=3$, then $R_p[\rho_{n,l}]$ does not depend on $n$ and so $R_p[\rho_{n,l,m}]=R_p[\rho_{n,l}]+R_p[Y_{l,m}]$, where the two summands are given by Eqs. \eqref{8} and \eqref{eq:angren} or \eqref{eq:ren_ang}, respectively.
\end{enumerate}
In particular, we have obtained the values
\begin{equation}
\label{eq:radcero}
R_{p}[\rho_{n,0,0}] = R_{p}[\rho_{n,0}] + R_{p}[Y_{0,0}]
\left\{
\begin{array}{ll}
 \simeq R_{p}[\rho_{n,0}]\to+\infty\,,\quad & p\in]0,3[\\
\\
=\ln(4\pi)-\frac12 \ln[4 \lambda^3C_B(\alpha,-1,3)] ,\quad & p=3 \\ 
\\
\simeq R_{p}[\rho_{n,0}]\to-\infty,\quad & p>3
\end{array}
\right.\;,
\end{equation}
for the total Rényi entropy of the state $(n\to\infty,0,0)$, and the values 
\begin{equation}
\label{eq:radone}
	R_{p}[\rho_{n,1,0}] = R_{p}[\rho_{n,1}] + R_{p}[Y_{1,0}]
\left\{
\begin{array}{ll}
\simeq R_{p}[\rho_{n,1}]\to+\infty\,,\quad & p\in]0,3[\\
\\
=\frac1{1-p}\ln[\frac{3^p(4\pi)^{1-p}}{2p+1}]-\frac12 \ln[4 \lambda^3C_B(\alpha,-1,3)] ,\quad & p=3 \\ 
\\
\simeq R_{p}[\rho_{n,1}]\to-\infty,\quad & p>3
\end{array}
\right.\;,
\end{equation}
for the total Rényi entropy of the state $(n\to\infty,1,0)$. Note that the radial entropies $R_{p}[\rho_{n,0}]$ and $R_{p}[\rho_{n,1}]$ in Eqs. \eqref{eq:radcero} and \eqref{eq:radone}, respectively, can be easily derived from \eqref{8} with $\beta = 1/2(1-p)$ and $\alpha = 1/2, 3/2$, respectively. From these results \eqref{eq:radcero} and \eqref{eq:radone} it can be easily shown that the Shannon entropy and the disequilibrium grow as $3/2\, ln\, n$ and $1/2\, ln\, n$, respectively.
\\

\section{Position-momentum Rényi and Shannon uncertainty sums}
\label{sec:5}
In this section we illustrate that our entropy results for the states at both extreme regions of the oscillator's energetic spectrum satisfy the known entropic uncertainty relations based on the Shannon entropy \cite{bialynicki1,rudnicki} and the Rényi entropy \cite{vignat,bialynicki2}. We begin by taking into account that, as already pointed out at the end of Section II, the quantum probability densities in the position and momentum spaces of our system are related through
\begin{equation}
\label{eq:rel_pos_mom}
\gamma_{n,l,m}(\vec{p}) = \frac{1}{\lambda^{3}}\rho_{n,l,m}\left(\frac{\vec{p}}{\lambda}\right),
\end{equation}
so that the Rényi entropy in momentum space can be obtained from the position entropy as
\begin{equation}
\label{eq:ren_mom}
R_{p}[\gamma_{n,l,m}] = R_{p}[\rho_{n,l,m}] + 3\ln\lambda, \quad p\neq 1.
\end{equation}
Then, for the $ns$-states (i.e., states with $l=m=0$) we have that the joint position-momentum Rényi-entropy-based uncertainty sum has the value 
\begin{equation}
\label{eq:ren_unc_sum}
R_{p}[\rho_{n,0,0}] + R_{q}[\gamma_{n,0,0}]  = \ln \left\{ [N_{n,0}(q)]^{\frac{1}{1-q}} [N_{n,0}(p)]^{\frac{1}{1-p}}\right\} + 2\ln(2\pi), \quad \frac{1}{p}+\frac{1}{q} =2, \quad \forall (n,0,0).
\end{equation}
(where we have taken into account Eqs. \eqref{eq:renyi4} and \eqref{eq:ren_mom}). In particular, this expression gives the value \begin{equation}
\label{eq:ren_unc_sum_0}
R_{p}[\rho_{0,0,0}] + R_{q}[\gamma_{0,0,0}]  = \ln \left[\left(\frac{\pi ^{\frac{1-p}{2}}}{p^{3/2}}\right)^{\frac{1}{1-p}} \left(\frac{\pi ^{\frac{1-q}{2}}}{q^{3/2}}\right)^{\frac{1}{1-q}}\right]+2 \ln \pi ,
\end{equation}
for the ground state $(n,l,m)= (0,0,0)$ of the harmonic oscillator, which saturates the Bialynicki-Birula-Zozor-Vignat Rényi-entropy-based  uncertainty relation \cite{vignat,bialynicki2}. Moreover, for $p \rightarrow 1$ and $q\rightarrow 1$, this expression gives the value
\begin{equation}
\label{eq:shan_unc_sum_0}
S[\rho_{0,0,0}] + S[\gamma_{0,0,0}]  = 3 (1+\ln \pi ) ,
\end{equation}
for the joint Shannon uncertainty sum of the oscillator ground-state, which saturates the celebrated Shannon-entropy uncertainty relation of Bilaynicki-Birula and Mycielski  \cite{bialynicki1}. Starting with Eqs. \eqref{eq:radcero} and operating in a similar way we can obtain the corresponding expressions for the position and momentum Rényi and Shannon entropies of the Rydberg oscillator-like states $(n\to\infty,0,0)$, which again verify the Rényi-entropy-based and Shannon-entropy-based uncertainty relations.
 
Moreover, let us now consider the oscillator states $(n,l,m)= (1,l,0)$. Then, one has that the joint Rényi-entropy-based uncertainty sum is
\begin{equation}
	R_{p}[\rho_{1,l,0}] + R_{q}[\gamma_{1,l,0}] = R_{p}[\rho_{1,l}] + R_{q}[\gamma_{1,l}] + R_{p}[Y_{l,0}] + R_{q}[Y_{l,0}]
\end{equation}
which, taking into account Eq. \eqref{eq:renyi4}, transforms into 
\begin{equation}
\label{eq:usum}
	R_{p}[\rho_{1,l,0}] + R_{q}[\gamma_{1,l,0}] = \ln \left\{ [N_{1,l}(p)]^{\frac{1}{1-p}} [N_{1,l}(q)]^{\frac{1}{1-q}}\right\} + R_{p}[Y_{l,0}] + R_{q}[Y_{l,0}] - 2\ln 2,
\end{equation}
where the radial integral can be shown (see the Appendix \ref{app}) to have the value
\begin{equation}
\label{eq:N1l}
N_{1,l}(p)=\frac{\Gamma(lp+\frac32)}{\Gamma(l+\frac52)^{p}}\frac{(2p)!}{p^{(l+2)p+\frac32}}L_{2p}^{(-(l+2)p-\frac32)} \left(-\left(l+\frac32\right)p\right),
 \end{equation}
In particular for states with $l=0$, since  $R_{p}[Y_{0,0}] = \ln(4\pi) $, one has from Eq. \eqref{eq:usum} that 
\begin{eqnarray}
\label{eq:ren_unc_sum1}
R_{p}[\rho_{1,0,0}] + R_{q}[\gamma_{1,0,0}]  &=& \ln \left\{ [N_{1,0}(p)]^{\frac{1}{1-p}} [N_{1,0}(q)]^{\frac{1}{1-q}}\right\} + 2\ln(2\pi),
\end{eqnarray}
where the integral $N_{1,0}(p)$ can be easily obtained from Eq. \eqref{eq:ren_unc_sum1}. Now, it is straightforward to check that this value verifies the Rényi-entropy-based uncertainty relation. Finally, let us point out that starting with Eqs. (\ref{8}), (\ref{eq:ren_mom}) and  (\ref{eq:ren_unc_sum}) and operating similarly we can readily see that the joint Rényi uncertainty sum of the Rydberg states $(n\to\infty,1,0)$ satisfy this entropic uncertainty relation as well.  \\

\section{Conclusions}

The harmonic systems are possibly the best studied finite systems in quantum physics since their wave equation can be exactly solved and because of their so many useful applications in science and technology. However, the knowledge of their information-theoretic measures is scarce and little known. Indeed, the spreading or spatial extension of the quantum-mechanical density of the isotropic harmonic oscillator has been examined by means of their central moments, particularly the second one (i.e., the variance) \cite{zozor}, almost up until now. However, their entropic measures (which are much more adequate to quantify the density of the oscillator-like states because they do not depend on any specific point of the system's region, contrary to what happens with the moments about the origin and the central moments) have been scarcely studied \cite{gadre1,yanez_1994,assche_1995,dehesa_1998,ghosh,dehesa_2001} and their determination is yet incomplete. This work has partially filled up this lack in the two following analytical ways.\\

We have determined the angular part of the basic entropic measures (Rényi, Tsallis, Shannon, disequilibrium) of the single-particle probability density which characterize the quantum states of ANY central potential. Then, we have computed the values of the total (i.e., angular+radial parts) values of these entropies for the highly-energetic oscillator-like states whose utility and multidirectional relevance is well-known. Finally we have performed the analytical calculation of the dominant term of the radial part in the Rydberg case, what is specially remarkable because it is a serious problem even numerically. Indeed, a naive use of quadratures for the numerical evaluation of the involved entropic functionals of the Laguerre polynomials which control the harmonic states is not convenient because the increasing number of integrable singularities spoil any attempt to achieve reasonable accuracy even for rather small values of $n$, since all the zeros of $L_n(x)$ belong to the interval of orthogonality. Moreover, we have illustrated that the analytic determination of the information entropies of the low energy oscillator-like states is much simpler. Finally, we have shown that the entropy results obtained for the joint position-momentum uncertainty sum at both extreme regions of the harmonic energetic spectrum satisfy the known Shannon-entropy-based and Rényi-entropy-based uncertainty relations. \\

\section*{Acknowledgments}
This work was partially supported by the Projects FQM-7276 and FQM-207 of the Junta de Andaluc\'ia and the MINECO-FEDER (Ministerio de Economia y Competitividad, and the European Regional Development Fund) grants FIS2014-54497P and FIS2014-59311P. The work of I. V. Toranzo was financed by the program FPU of the Spanish Ministerio de Educación. 
\appendix
\section{Derivation of Equation \eqref{eq:N1l}}
\label{app}
 To obtain Eq.\eqref{eq:N1l} we start with the well-known \cite{nist} Euler's integral of the second kind $\Gamma(z)=\int_0^\infty e^{-t}\, t^{z-1}\,dt,\,\,\text{Re}(z)>0$, which can be easily extended as
\[
\int_0^\infty x^{\nu-1}\,e^{-\mu (x+a)}\,dx= \Gamma(\nu)\frac{e^{-\mu a}}{\mu^\nu},\quad \mu>0,\,\nu>0,\,a\in\mathbb  R.
\]
Now, the $n$th-derivative with respect to $\mu$ in this expression and taking into account the Rodrigues' formula of the Laguerre polynomials allow us to have that
\begin{equation}
\label{eq:lag_integ}
\int_0^\infty (a+x)^n\,x^{\nu-1}\,e^{-\mu x}\,dx= (-1)^n\,\Gamma(\nu)\cdot e^{\mu a}\frac{d^n}{d \mu^n}\left({e^{-\mu a}}{\mu^{-\nu}}\right)=(-1)^n\,\frac{\Gamma(\nu) n!}{\mu^{n+\nu}} L_n^{(-n-\nu)}(a \mu),
\end{equation}
which can be rewritten as  
\begin{equation}
L_n^{(-n-\nu)}(x)=\frac{(-1)^n}{ n!\,\Gamma(\nu)}\int_0^\infty (x+y)^n\,y^{\nu-1}\,e^{-y}\,dy,
\end{equation}
which gives an integral representation for the varying Laguerre polynomials with a negative parameter, what is interesting \textit{per se} in the field of orthogonal polynomials. Finally, for the particular case $2p\in\mathbb N$, we can insert (\ref{eq:lag_integ}) into (\ref{eq:c1.2}) to obtain the wanted expression
\begin{equation}
N_{1,l}(p)=\frac{\Gamma(lp+\frac32)}{\Gamma(l+\frac52)^{p}}\frac{(2p)!}{p^{(l+2)p+\frac32}}L_{2p}^{(-(l+2)p-\frac32)} \left(-\left(l+\frac32\right)p\right),
 \end{equation}
 which holds for the states with $(n=1, l=0)$ and $(n=1, l=1)$.

\end{document}